# $C_{60}$ as a probe for astrophysical environments

A.C. Brieva[1,*], R. Gredel[2], C. Jäger[1], F. Huisken[1] and T. Henning[2]

[1] Laboratory Astrophysics Group of the Max Planck Institute for Astronomy at the Friedrich Schiller University Jena, Institute of Solid State Physics, Helmholtzweg 3, 07743 Jena, Germany.

[2] Max Planck Institute for Astronomy (MPIA), Königstuhl 17, 69117 Heidelberg, Germany

[*] aab01@alumni.aber.ac.uk

ABSTRACT

The $C_{60}$ molecule has been recently detected in a wide range of astrophysical environments through its four active intramolecular vibrational modes ($T_{1u}$) near 18.9 µm, 17.4 µm, 8.5 µm, and 7.0 µm. The strengths of the mid-infrared emission bands have been used to infer astrophysical conditions in the fullerene-rich regions. Widely varying values of the relative intrinsic strengths (RIS) of these four bands are reported in laboratory and theoretical papers, which impedes the derivation of the excitation mechanism of $C_{60}$ in the astrophysical sources. The spectroscopic analysis of the $C_{60}$ samples produced with our method delivers highly reproducible RIS values of 100, 25 ± 1, 26 ± 1 and 40 ± 4. A comparison of the inferred $C_{60}$ emission band strengths with the astrophysical data shows that the observed strengths cannot be explained in terms of fluorescent or thermal emission alone. The large range in the observed 17.4 µm/18.9 µm emission ratios indicates that either the emission bands contain significant contributions from emitters other than $C_{60}$, or that the population distribution among the $C_{60}$ vibrational modes is affected by physical processes other than thermal or UV excitation, such as chemo-luminescence from nascent $C_{60}$ or possibly, Poincaré fluorescence resulting from an inverse internal energy conversion. We have carefully analyzed the effect of the weakly-active fundamental modes and second order modes in the mid-infrared spectrum of $C_{60}$ and propose that neutral $C_{60}$ is the carrier of the unidentified emission band at 6.49 µm which has been observed in fullerene-rich environments.

1. INTRODUCTION

The discovery of $C_{60}$ in the laboratory (Kroto et al. 1985) three decades ago triggered an intensive search of fullerenes in astrophysical environments. The presence of the neutral $C_{60}$ was proposed previously (e.g. review by Herbig 2000), yet the first firm detection of $C_{60}$ has been presented by Cami et al. (2010) in the planetary nebula Tc-1. Subsequently, fullerenes have been identified in a wide range of astrophysical environments, including the reflection nebulae NGC 7023 and NGC 2023 (Sellgren et al. 2010), the proto-planetary nebula IRAS01005+7910 (Zhang & Kwok 2011), the circumstellar envelopes of post-asymptotic giant branch stars and molecular outflows from young stellar objects (Gielen et al. 2011; Roberts et al. 2012), and in a large number of planetary nebulae in the Milky Way and the Magellanic Clouds (García-Hernández et al. 2012b; Otsuka et al. 2014; Sloan et al. 2014). Mid-infrared (MIR) emission from $C_{60}^+$ has been reported in the reflection nebula NGC 7023, by Berne et al. (2013), and the confirmation of the longstanding hypothesis that interstellar $C_{60}^+$ is the carrier of two diffuse interstellar bands at 9577 Å and 9632 Å (Foing & Ehrenfreund 1994) by laboratory measurements has recently been claimed by Campbell et al. (2015) and Walker et al. (2015).
While $C_{60}^+$ in the diffuse interstellar medium has been firmly identified via electronic absorptions at 9577 Å and 9632 Å (Ehrenfreund & Foing 2015), neutral $C_{60}$ is detected via its four active intramolecular vibrational modes [1]: $T_{1u}(4)$, $T_{1u}(3)$, $T_{1u}(2)$, and $T_{1u}(1)$, at wavelengths near 7.0 µm, 8.5 µm, 17.4 µm, and 18.9 µm, respectively. The strengths of the mid-infrared emission bands have been used to infer the excitation mechanism of $C_{60}$. Not surprisingly and in analogy to simpler molecules such as $H_2$ in photon-dominated regions, the observational data are interpreted either in terms of a thermal excitation scenario implying emission from warm, solid-state $C_{60}$ (Cami et al. 2010; García-Hernández et al. 2012b; Roberts et al. 2012), or by UV excitation of gas-phase $C_{60}$ (e.g. Sellgren et al. 2010).

The study of astrophysical fullerenes is also interesting because it can shed light on the condensation processes going on in carbon-rich regions and contribute to elucidate the formation pathways of the fullerenes, which are still under debate. In analogy to the laboratory condensation of hydrocarbons from gas phase, fullerene formation might be enhanced in hydrogen-poor regions (e.g. Jäger et al. 2009), although photochemical processing of hydrogenated amorphous carbon (HAC) particles has been proposed as a potential formation scenario as well (Micelotta et al. 2012). The amount of carbon involved in the astrophysical fullerenes is also noteworthy: a fraction larger than 1% of the available carbon has been estimated in some astrophysical objects (Cami et al. 2010). The stability of $C_{60}$ under harsh conditions (Castellanos et al. 2014; Cataldo et al. 2009; Ehrenfreund & Foing 2010) might convert this material into a significant component of the interstellar medium (Bernard-Salas et al. 2014).

The interpretation of the astrophysical emission of $C_{60}$ requires accurate laboratory measurements on both the transition wavelengths of the $T_{1u}$ modes as well as the relative intrinsic strengths (RIS) of the four bands. Ideally, these data are obtained in emission and under astrophysically relevant conditions. However, infrared emission spectra are generally obtained in a range of pressures and temperatures which are not comparable to the astrophysical environments (Frum et al. 1991; Nemes et al. 1994). Therefore, the RIS of $C_{60}$ are obtained from laboratory spectra in absorption. For the interpretation of the astrophysical emission, RIS values are converted into expected emission band ratios using the Einstein relations and assuming either a thermal distribution among the vibrational modes, or a distribution resulting from a molecular fluorescence process with a micro-canonical temperature, which is similar to the generally accepted excitation mechanism of polycyclic aromatic hydrocarbons (Léger et al. 1989; Sellgren 1984).

The various laboratory studies published on $C_{60}$ in the solid phase at room temperature (Chase et al. 1992; Fu et al. 1992; Iglesias-Groth et al. 2011; Martin et al. 1993; Onoe & Takeuchi 1996; Winkler et al. 1994) are largely inconsistent with each other (Bernard-Salas et al. 2012; Castellanos et al. 2014; Tielens 2013; Zhang & Kwok 2013), and even larger discrepancies occur when theoretical results are taken into account (e.g. Zhang & Kwok 2013 and references therein). The existing inconsistencies motivated us to present new data on the RIS values of the four infrared-active modes. In Section 2, we discuss our experimental method for preparing the samples. In Section 3 we discuss the validity of our results at temperatures from above 260 K to the sublimation temperature, and their applicability for the interpretation of gas-phase spectra. Our analysis of the spectroscopic data includes the contribution of the weak bands in the MIR. The RIS values presented here are applied to the interpretation of the available astrophysical data in Section 4.

## 2. EXPERIMENTAL

A detailed description of the experiments is given in Appendix A. In Figure 1(a) we show a schematic of the crucible used in the preparation of the samples. $C_{60}$ was evaporated by laser desorption and deposited on potassium bromide substrates. We produced the 36 samples analyzed in the present work using the same laser flux and varying the deposition time in order to obtain samples of different thicknesses.

---

[1] We employ the Mulliken symbols commonly used in the literature to denote vibrational modes. We refer to the bands assigned to the $T_{1u}(i)$ (i =1, 2, 3, 4) modes as $T_{1u}(i)$ bands.

In previous experiments with the same setup but using different values of laser flux, we observed that yellow-orange spots of $C_{60}$ resulted from the deposition with power densities just above the minimum necessary for the evaporation of the original $C_{60}$ powder, while black spots were produced when higher values were used (Figure 1(b)). We determine that the threshold for the power density between both situations was around 4 W cm$^{-2}$. Importantly, MIR spectra of the spots deposited using power densities above this threshold showed a high variability in the RIS of the $T_{1u}$ bands, probably due to the formation of amorphous carbon (Section 3.2).

All the samples used in the analysis were produced with a power density below the threshold. The only morphological requirement for the samples was having a thickness within the optimum range for MIR transmission spectroscopy, but allowing large differences from sample to sample and also at different points on each sample.

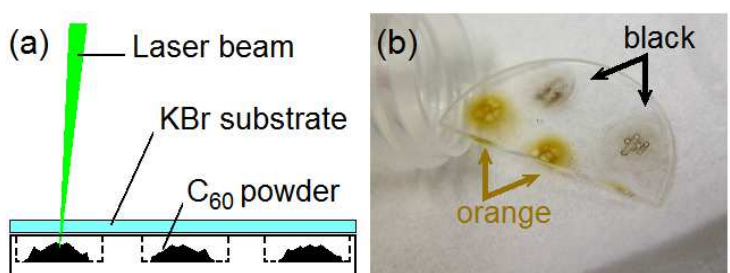

*Figure 1. a) Schematic of the laser desorption setup b) Semicircular KBr substrate after deposition of $C_{60}$ using two different values of laser power. Black and orange arrows point to the condensates obtained using average power densities above and bellow the threshold of 4 W cm$^{-2}$ (i.e. 100 MW cm$^{-2}$ during the pulse).*

3. RESULTS AND DISCUSSION

In Section 3.1 we discuss our analysis procedure considering the small bands in the MIR spectrum of the $C_{60}$. In Section 3.2 we present our experimental results on the intrinsic strength of the $T_{1u}$ bands. Finally, in Section 3.3 we propose that neutral $C_{60}$ is the carrier of the unidentified band at 6.49 μm observed in fullerene-rich astrophysical objects.

We justify the use of solid $C_{60}$ for the experiments and discuss the temperature range of validity of our results in Appendix B.

3.1 Experimental data and spectroscopic analysis

$T_{1u}(1)$, $T_{1u}(2)$, and $T_{1u}(3)$ bands appear isolated in the MIR and can be described by single curves. The case of the $T_{1u}(4)$ band is more complex because of the presence of a large number of small bands in the 7 μm region. We discuss the nature of those bands in Section 3.1.3. Figure 2 illustrates the typical shape of the spectra in the region containing the $T_{1u}(4)$ band. In Figure 2(a) we present a spectrum of one of the samples showing the general structure of this region, which can be modeled as a superposition of five components. The components used for the fitting are shown in the synthetic curve in Figure 2(b). The low-frequency side of this structure is dominated by the $T_{1u}(4)$ mode at 1429.3 cm$^{-1}$ (7.00 μm) with a shoulder at 1424 cm$^{-1}$ (7.02 μm). An intermediate plateau would be formed by the contributions of two components at 1444 and 1451 cm$^{-1}$ (~ 6.9 μm), and a broad band centered at 1462 cm$^{-1}$ (~ 6.8 μm) would be at the higher frequency side.

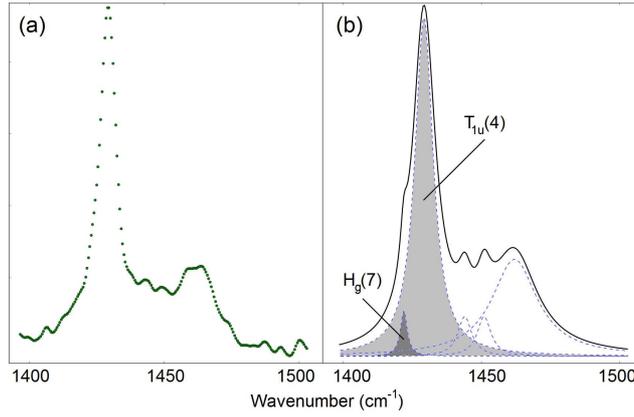

*Figure 2. Shape of the spectra at the 7 μm region containing the $T_{1u}(4)$ band. a) Spectrum of one of the samples, b) Synthetic spectrum reproducing the general shape of the spectra at the 7 μm region, which consists of five components (dashed curves). Due to its proximity to $T_{1u}(4)$, the band at 1424 cm$^{-1}$ appears as a shoulder. Parameters of the components are listed in Table 1.*

Since the main purpose of the fitting of the structure in the 7 μm region is to determine the parameters of the $T_{1u}(4)$ band, some small features which appear at different positions in every spectrum have not been considered for the fitting. On the other hand we have regarded the peaks at 1444 and 1451 cm$^{-1}$ (~ 6.9 μm) because they distinctly appear in a number of the spectra recorded and, together with the broad peak at 1462 cm$^{-1}$ (~ 6.8 μm), reproduce the structure common to all of them. The case of the small band responsible for the shoulder at 1424 cm$^{-1}$ is discussed in the following sections.

3.1.1 Curve fitting

The purpose of the curve fitting is to determine the intrinsic strength (integrated absorptivity) of each of the four $T_{1u}$ bands, which is given by the areas below the fitting curve.

We have found that all the bands can be well described by Lorentzian functions, as expected from a quasi-molecular solid such as solid $C_{60}$ (Section 3.1.3). The Gaussian character introduced by the instrumental broadening can be taken into account, if necessary, by carrying out the fit with Voigt profiles. With our instrumental resolution (better than 0.2 cm$^{-1}$) and the width of the bands evaluated here, Voigt curves and Lorentzian profiles do not yield significantly different results. There are indications of a weak scattering in some of the spectra, so a small correction has been introduced in the fitting process. For the optimization of the fitting, a non-linear least squares algorithm was used.

In Figure 3, we present the fitting of one of the spectra of the 36 samples analyzed in the present work. The fitting of $T_{1u}(1)$, $T_{1u}(2)$, and $T_{1u}(3)$ bands does not present major problems and most of the residual seems to be due to noise. In Figure 3(d) we show the fitting of the structure that contains the $T_{1u}(4)$ band. All the spectra recorded present this general shape, although the small components at 1444 and 1451 cm$^{-1}$ (~ 6.9 μm) can only be clearly located in some of them. The largest contribution to the low-frequency side of this structure is assigned to the $T_{1u}(4)$ band. An intermediate plateau is formed by the contribution of the 1444 and 1451 cm$^{-1}$ peaks. Finally, we used a broad peak centered at 1462 cm$^{-1}$ (~ 6.8 μm) to fit the higher-frequency side although, in some of the spectra, it seems to be formed by two components at 1457.5 and 1463.5 cm$^{-1}$. Due to its proximity to $T_{1u}(4)$ at 7.00 μm, the band at 1424 cm$^{-1}$ (7.02 μm) appears as a shoulder, however its position is well known because it corresponds to the Raman-active mode $H_g(7)$, which has already been reported to be weakly active in the MIR (Martin et al. 1994). The fitting in Figure 3(d) also includes the small band near 6.5 μm discussed in Section 3.3. The parameters of the bands obtained during the fitting process and the possible assignments for the small bands are listed in Table 1.

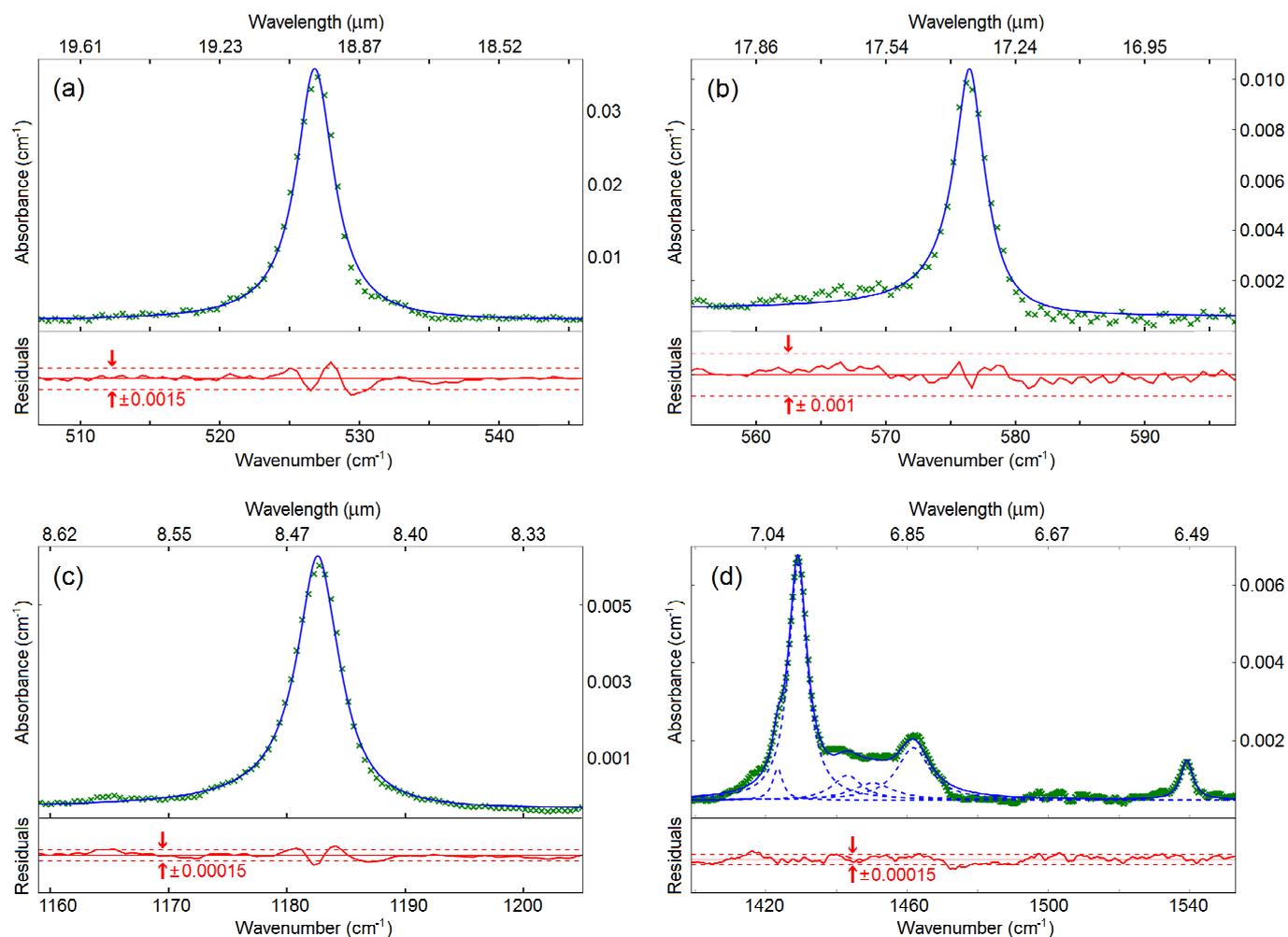

*Figure 3. Fit result of one of the samples used in the evaluation of the relative intrinsic strengths of the main $C_{60}$ MIR bands. In a), b), c) and d) we show the $T_{1u}(1)$, $T_{1u}(2)$, $T_{1u}(3)$ and $T_{1u}(4)$ bands respectively, where crosses represent experimental points and solid lines are used for the fitting curves. In d) the individual contributions to the spectrum in the 7 μm region discussed in the text are plotted as dashed curves. The fitting shown in d) also includes the small band at 1539.4 $cm^{-1}$ (6.496 μm).*

3.1.2 Sources of error and contaminants

In the range between 450 and 3500 $cm^{-1}$ (22.2 – 2.9 μm), we have removed the continuum of every spectrum with a cubic spline using anchor points free of absorption features. The background used for the fitting of each band was computed by averaging the experimental values at both sides of the peaks, and incorporates the effect of a weak scattering observed in some of the samples. Backgrounds within three standard deviations of the mean value were then used to estimate the random error in the parameters obtained from the fitting.

We did not find indications in the MIR of the presence of other fullerenes, such as $C_{70}$. Neither did we observe any photo-polymerization (Rao et al. 1993). Spectra of every substrate were recorded prior to deposition to check that no features due to contaminants were close to the positions of the $T_{1u}$ bands of the $C_{60}$.

To double-check the absence of contaminants we carried out Raman experiments. Figure 4 compares the VIS Raman spectrum of a typical $C_{60}$ film on KBr with the spectrum of the original $C_{60}$ powder used to prepare the $C_{60}$ films by laser desorption. Raman spectroscopy can be used for a compositional and structural analysis of carbonaceous materials, since impurities such as amorphous carbon, polymerized or oxidized $C_{60}$, that could be produced during laser desorption of the $C_{60}$ powder, would be clearly identifiable in the spectrum. According to Dong et al. (1993) and Rao, et al. (1993), $C_{60}$ exhibits 10 Raman active intramolecular modes (2 $A_g$ + 8 $H_g$). All of them can be observed in the spectrum of the original $C_{60}$ (Fig.4, lower curve) with the exception of the two very weak bands at 1100 and 1251 cm$^{-1}$ (9.09 and 7.99 μm). Raman spectra of the produced films do show all the bands of $C_{60}$ (Fig.4, upper curve). The very sensitive G and D lines at 1600 and 1350 cm$^{-1}$ (6.25 and 7.41 μm), typical of amorphous or hydrogenated amorphous carbon (Nemanich et al. 1988) are not visible. Furthermore, no hints of photo-polymerization or oxidation products that give rise to the formation of many additional lines in the first order Raman spectrum can be detected (Rao, et al. 1993). The Raman analysis clearly confirms the purity of the produced $C_{60}$ film.

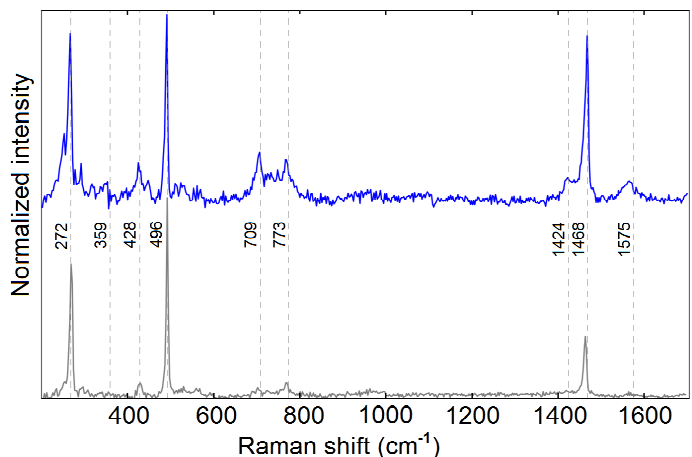

*Figure 4. VIS Raman spectrum of a $C_{60}$ film on KBr prepared by the experimental procedure described in this paper (upper curve) compared to the spectrum of the original $C_{60}$ powder (lower curve). The spectrum was taken at room temperature using a He-Ne laser as excitation source. The frequencies labeled correspond to fundamental frequencies of the $A_g$ and $H_g$ modes.*

3.1.3 Nature of the weak bands in the MIR spectrum of solid $C_{60}$

Due to the high symmetry of the molecule, $C_{60}$ has a very simple vibrational spectrum where most of the vibrational modes were expected to be silent. However, in addition to the four intense bands assigned to the $T_{1u}$ modes, a large number of small bands were found (Kraetschmer et al. 1990b). These weak bands are particularly noticeable below 8 μm, and were regarded with suspicion by the early researchers, who had to prepare and purify the $C_{60}$ by themselves following the method described by Kraetschmer et al. (1990a). They often attributed the appearance of the small bands to the presence of impurities, such as other fullerenes, rests of solvents, or functional groups (Chase et al. 1992; Haufler et al. 1990; Kraetschmer et al. 1990b). Later, when increasingly purer fullerenes were available, it became clear that the carrier of the weak bands was indeed $C_{60}$.

The nature of the weak bands in the MIR spectrum of solid $C_{60}$ was discussed by Martin et al. (1994) and Wang et al. (1993) while the Raman spectrum was studied by Dong et al. (1993). The remarkable sharpness of the weak bands (indicating small dispersion of the phonon modes) proves that solid $C_{60}$ is an almost ideal molecular crystal with very weak intermolecular interaction. These authors initially considered that the probable cause of the activation of the silent modes could be the breaking of the symmetry caused by the inclusion of $^{13}C$ atoms in the molecule. However, the same researchers later discarded the hypothesis of the isotopic activation (Martin et al. 1995). Mechanical and electrical anharmonicities were proposed by Fabian (1996) as possible mechanisms for activating the silent modes. Anyhow, the strongest of these bands, near 6.49 μm (1539 cm$^{-1}$), has been observed in the gas phase (Frum et al. 1991; Nemes et al. 1994) and in $C_{60}$ isolated in neon matrices (Kern et al. 2013), therefore, its presence should not be due to intermolecular interaction. We will discuss further on this band which also appears in astronomical observations.

It is clear that most of the small features in the vibrational spectrum must be attributed to combination modes. The uncertainty in the frequencies of the silent fundamental modes (explored by inelastic neutron scattering) and their proximity to each other in the 7 μm region, makes it difficult to reach a definitive assignment. We have listed in Table 1 the most likely candidates for the bands discussed in this work, using the values given by Menendez & Page (2000) for the frequencies of the fundamentals.

Regardless of the open questions on the nature and theoretical assignment of these bands, two facts in the literature cited above are important for the present study: the quasi-molecular nature of solid $C_{60}$ at room temperature, and the relevance of the contribution of the weak bands observed in the evaluation of the RIS, since the $T_{1u}(4)$ band is embedded in a structure formed by a number of weaker components.

3.2 Relative intrinsic strengths obtained

The RIS values obtained are listed in Table 2. They show small dispersion in spite of the large differences between samples (in morphology and thickness, as exposed in the experimental section). Taking into account that the physical state and morphology of the cosmic $C_{60}$ is unknown, we believe that the astrophysical significance of the present study relies on the reproducibility of our spectroscopic data (in particular on the stability of the RIS ratios), regardless of the morphological differences between samples.

In Table 2, we propose three possible sets of values to be used in the excitation models to predict the astrophysical emission, with three different relative strengths for the 7 μm line. First with the contribution of the $T_{1u}(4)$ mode alone, second including the contribution of the $H_g(7)$ mode at 1424 cm$^{-1}$ (7.02 μm) discussed above, and third considering the whole structure at the 7 μm region, which included the contribution of the small bands listed in Table 1. Uncertainties in the values are given by the statistical fluctuations from the fitting of the experimental spectra, which are larger when more contributions are considered. It is up to the astronomers to make the choice between them according to the resolution of the observational spectra available, and the hypothesis about the activation of the infrared modes in the astrophysical sources. The contribution of the $H_g(7)$ mode to the astrophysical emission is an open issue since, as discussed above, the causes of the activation of the infrared silent modes are not entirely clear (Section 3.1.3). However, since the $H_g(7)$ mode has been found to be infrared active in the laboratory (Martin et al. 1994), and its proximity to the $T_{1u}(4)$ band makes it impossible to separate both contributions with the astronomical instrumentation available, we propose the use of the RIS of 100: 25: 26: 40.

To explain the large discrepancies between the RIS reported in the experimental studies published so far, we consider that the most probable cause would be the interaction of $C_{60}$ with some other materials, such as amorphous carbon generated by partial decomposition of $C_{60}$, or the light scattering occurring in non-homogeneous solid samples (Christiansen effect [2]). Regarding the first, we observed in previous experiments during the preparation of our set-up, that an excess of power density used in the laser desorption produces samples with high RIS variability (Section 2). Formation of amorphous carbon from the fragmentation of the molecules has been reported by Canulescu et al. (2011) in matrix-assisted laser desorption of $C_{60}$ above a power density comparable (regarding the experimental differences) to our threshold value. Also, our current research indicates that the RIS values are very sensitive to the interaction with other forms of carbon (A.C. Brieva et al. 2016, in preparation). Considering the impact of the Christiansen effect, we noticed that some of the spectra published are affected by relatively strong scattering. For an accurate estimation of the RIS from these spectra, extra parameters would be required in the modeling (Franz et al. 2008; Scollo et al. 2013).

For reference, our average experimental positions and full widths at half maximum (FWHM) of the $T_{1u}$ bands are listed in Table 3.

---

[2] The alteration of the absorption profiles of powders embedded in a matrix is named after Christian Christiansen who described this effect in 1884.

3.3 Carrier of the unidentified emission band at 6.49 µm

From the study of the weak bands of $C_{60}$ in the MIR, although not affecting the RIS values, we notice that the strongest of them, at $1539.4 \pm 0.2$ cm$^{-1}$ ($6.496 \pm 0.002$ µm), is visible in all of our spectra (Figure 3(d)) and in almost every spectrum published (Section 3.1.3). Since the feature is close to the expected position of one of the $C_{60}^+$ bands discussed in Section 4, we notice that in our experimental setup there were no sources of ionization, and therefore we can rule out the presence of $C_{60}^+$ in our samples. The same can be inferred from the experimental information of the data published. This band must be assigned to one or perhaps two (considering its strength) second order modes. In Table 1, we give some possible assignments based on the literature. The position of this band matches the little peak reported at 6.49 µm in some fullerene-rich astrophysical objects, such as the planetary nebula Tc-1 and SMP-LMC-56 (Bernard-Salas et al. 2012) and the post-AGB star IRAS 06338 (Gielen et al. 2011). It has been proposed that activation of silent modes in astrophysical $C_{60}$ could be due to the formation of fullerene-metal complexes (Dunk et al. 2013). If so, the detection of these bands would represent a new insight into the chemistry of the astrophysical fullerene-rich environments. In the case of this band, however, regarding the fact that it appears in almost every MIR spectra published, we consider that neutral $C_{60}$ is the most probable carrier.

4. DETECTION OF FULLERENES IN ASTROPHYSICAL ENVIRONMENTS

The discovery of $C_{60}$ in the planetary nebula Tc-1 by Cami et al. (2010) via its four active intramolecular vibrational modes $T_{1u}(4)$, $T_{1u}(3)$, $T_{1u}(2)$, and $T_{1u}(1)$, at wavelengths near 7.0 µm, 8.5 µm, 17.4 µm, and 18.95 µm, respectively, has triggered a wealth of studies of fullerenes in a wide range of astrophysical environments. Detections have been reported in the reflection nebula NGC 7023 (Sellgren et al. 2010), in the proto-planetary nebula IRAS 01005+7910 (Zhang & Kwok 2011), the oxygen-rich post-AGB stars HD 52961 and IRAS 06338 (Gielen et al. 2011), the Orion Nebula (Rubin et al. 2011), the R Coronae Borealis stars CPD-56 8032 (Clayton 2011) and DY Cen and V854 (Garcia-Hernandez 2011b), the peculiar binary XX Oph (Evans et al. 2012), in the young stellar objects (YSO) ISOGAL-PJ174639.6-284126, SSTGC372630 and the YSO candidate 2MASSJ06314796+0419381 (Roberts et al. 2012), in the Herbig Ae/Be star HD 97300 (Roberts et al. 2012), and in several planetary nebulae in the Galaxy and the Magellanic Clouds (Garcia-Hernandez et al. 2012b, Otsuka et al. 2014, Sloan et al. 2014).

The observed flux in the four $T_{1u}(i)$ bands has been used to infer the excitation scenario of astrophysical $C_{60}$. The analysis was pioneered by Cami et al. (2010), who used the RIS values of Martin et al. (1993) and Fabian et al. (1996) to infer the population distribution among the vibrational states. As noted by the authors, their analysis is impeded by the large spread of $C_{60}$ RIS values that are available from the literature, and by contributions to the emission from species other than $C_{60}$ in the four $T_{1u}(i)$ bands. In the following, we revisit the available data on the $C_{60}$ emission bands and the conclusions drawn about the relevant excitation mechanisms. In most cases, the $C_{60}$ emission has been explained in terms of thermal excitation, while UV excitation has been ruled out. We show that, in several cases, the conclusions have been misled by an inconsistent use of RIS values. We also show that some of the reported fluxes from observational data in the four $T_{1u}(i)$ bands suffer from large uncertainties, and are not sufficiently robust to infer physical conditions in astrophysical environments.

4.1 Thermal excitation of $C_{60}$

In their detailed analysis, Cami et al. (2010) explained that the emission detected in Tc-1 arise from thermally excited $C_{60}$ at temperatures around 330 K. The authors were forced to disregard the observed flux at 7 µm because it is dominated by emission from [ArII] at 7.03 µm. Thermal excitation scenarios were also invoked to explain the $C_{60}$ observations in the four planetary nebulae M1-12, M1-20, K3-54, and SMP SMC 16 (García-Hernández 2010b), in the proto-planetary nebula IRAS 01005+7910 (Zhang & Kwok 2011), and in the three YSOs studied by Roberts et al. (2012). All groups that followed the method of Cami et al. (2010) inferred excitation temperature ranges from about 460 K for IRAS

01005+7910, 320 - 680 K for M1-12, M1-20, K3-54, and SMP SMC 16, and 400 - 540 K in the three YSOs. The observed emission in the post-AGB stars HD 52961 and IRAS 06338 studied by Gielen et al. (2011) indicates excitation temperatures of about 150 K, again following Cami et al. (2010). The authors point out that the emission toward IRAS 06338 is dominated by high-excitation emission bands due to $CO_2$, which contaminate the 17.4 µm and 18.95 µm $C_{60}$ emission bands. Thermal excitation at 520 K were invoked by Evans et al. (2012) to model the observed emission bands at 17.25 µm and 18.95 µm towards XX Oph, using the $C_{60}$ RIS values of Mitzner & Campbell (1995). The emission in both bands is also consistent with excitations by 10 eV photons, but Evans et al. (2012) note for fluorescence mechanism, that emission in the 8.5 µm band is expected as well but not detected.

4.2 UV excitation of $C_{60}$

A molecular fluorescent mechanism was pioneered by Sellgren et al. (1984) to explain color temperatures of very small grains in excess of 1000 K in astrophysical environments, where the grain equilibrium temperature is several 10 K only. The concept to describe the UV excitation of large molecules in terms of a vibrational temperature has been firmly established by Léger et al. (1989) and d'Hendecourt et al. (1989) and is now commonly adopted to explain the infrared emission of interstellar polycyclic aromatic hydrocarbon molecules (PAH). The photo-excitation of $C_{60}$ has been extensively studied and is well understood (e.g. Stepanov et al. 2002, among others). The absorption of a 1 - 14 eV photons is followed by quasi-immediate transfer to adjacent electronic levels, and very fast internal energy conversion with a final transfer of the energy to vibrational states.

UV excitations have been invoked by Sellgren et al. (2010) to explain the origin of the $C_{60}$ emission detected in the reflection nebulae NGC 7023 and NGC 2023. Using the $C_{60}$ band strengths of Choi et al. (2000) and a detailed micro-canonical treatment, Sellgren et al. (2010) predicted a ratio of the 17.4 µm/18.95 µm emission bands in the range of 0.28 - 0.38, basically insensitive to the UV photon energy. The observed 17.4 µm/18.95 µm ratios in NGC 7023 and NGC 2023 range from 0.33 to 0.66. Based on the imaging spectroscopy performed on NGC 7023, Sellgren et al. (2010) conclude that the 17.4 µm emission band contains significant contributions from PAH. In NGC 7023, the ratio depends on the spectral resolution of the observations, indicating that unresolved blends affect the low-resolution data. The modeled 7.0 µm/18.95 µm and 8.5 µm/18.9 µm ratios, on the other hand, increase significantly with the photon energy. The observed 7.0 µm/18.95 µm ratio in NGC 7023 is best explained in terms of excitation by 10 eV photons, while very low photon energies (< 5eV) would explain the emission in NGC 2023.

In their analysis of the $C_{60}$ emission from three YSOs, Roberts et al. (2012) ruled out an UV excitation scenario based on the observed 17.4 µm/18.95 µm flux ratios of 0.5 - 0.6, which are about a factor of two higher than the ratios modeled by Sellgren et al. (2010). However, Roberts et al. (2012) were misled by an inconsistent use of band strengths when comparing thermal and UV excitation of $C_{60}$. The modeled 17.4 µm/18.95 µm flux ratio depends critically on the adopted $C_{60}$ band strengths and the authors use the band strengths from Cami et al. (2010) and Sellgren et al. (2010), which differ by more than a factor of two, for fluorescent and thermal scenarios respectively. Indeed the modeled 17.4 µm/18.95 µm flux ratio for thermal and UV excitation yields very similar results when the same set of $C_{60}$ band strengths is used.

The inconsistent use of $C_{60}$ RIS also impedes the analysis of García-Hernandez et al. (2012b), who detected $C_{60}$ in a larger sample of PNe in the Galaxy and in the Magellanic Clouds. After the subtraction of potential contributions from $C_{70}$ to the observed 17.5 µm and 18.9 µm bands, and from [ArII] to the 7.0 µm emission band, García-Hernandez et al. (2012b) concluded that the observed $C_{60}$ emission in their sample of 11 $C_{60}$-emitting PNe cannot be explained in terms of photo-excitation from the central star. The authors followed the approach of Cami et al. (2010) and Sellgren et al. (2010), but did not account for the difference in the $C_{60}$ band strengths used in the latter two studies.

### 4.3 $C_{60}$ as a probe for physical conditions in astrophysical environments

In the following, we use our RIS values to interpret the existing astrophysical data on $C_{60}$. In Figs. 5 and 6, we present the observed 8.5 μm/18.95 μm vs. the 17.4 μm/18.95 μm flux ratios, and the 7.0μm/18.95 μm vs. the 17.4 μm/18.95 μm flux ratios taken from Sellgren et al. (2010), García-Hernandez et al. (2012b), Bernard-Salas et al. (2012), Otsuka et al. (2014), and Sloan et al. (2014). Our RIS values are converted into emission line ratios using the Einstein relations and assuming either thermal emission with a Boltzmann distribution at the equilibrium temperature, or emission at a micro-canonical temperature $T_m$ resulting from the molecular fluorescence mechanism described by Tielens (2013), that relates the UV photon energy to a micro-canonical temperature $T_m$. The modeled 8.5 μm/18.95 μm vs. 17.4 μm/18.95 μm and 7.0μm/18.95 μm vs. 17.4 μm/18.95 μm flux ratios are represented in Figs. 5 and 6. The solid black curve represents the expected ratios for a thermal excitation with temperatures ranging from 300 K to 1000 K. The blue points represent the expected ratios for UV excitations, for UV photon energies ranging from 1eV to 10 eV.

Using our RIS values, we are able to model the observed $C_{60}$ emission band ratios for the planetary nebulae in the Small Magellanic Cloud SMC-13, SMC-15, SMC-16, and SMC-24, and the Large Magellanic Cloud PN LMC-56 as presented by Sloan et al. (2014). For a thermal excitation scenario, we infer temperatures of 350 K - 550 K for SMC-13 and SMC-15. Alternatively, the observations are consistent with UV excitation by photons in range of 1 - 10 eV. For SMC-16, SMC-24, and LMC-56, the observed band ratios would indicate rather low thermal excitation temperatures (below 300 K), or excitations by UV photons with energies below 1 eV.

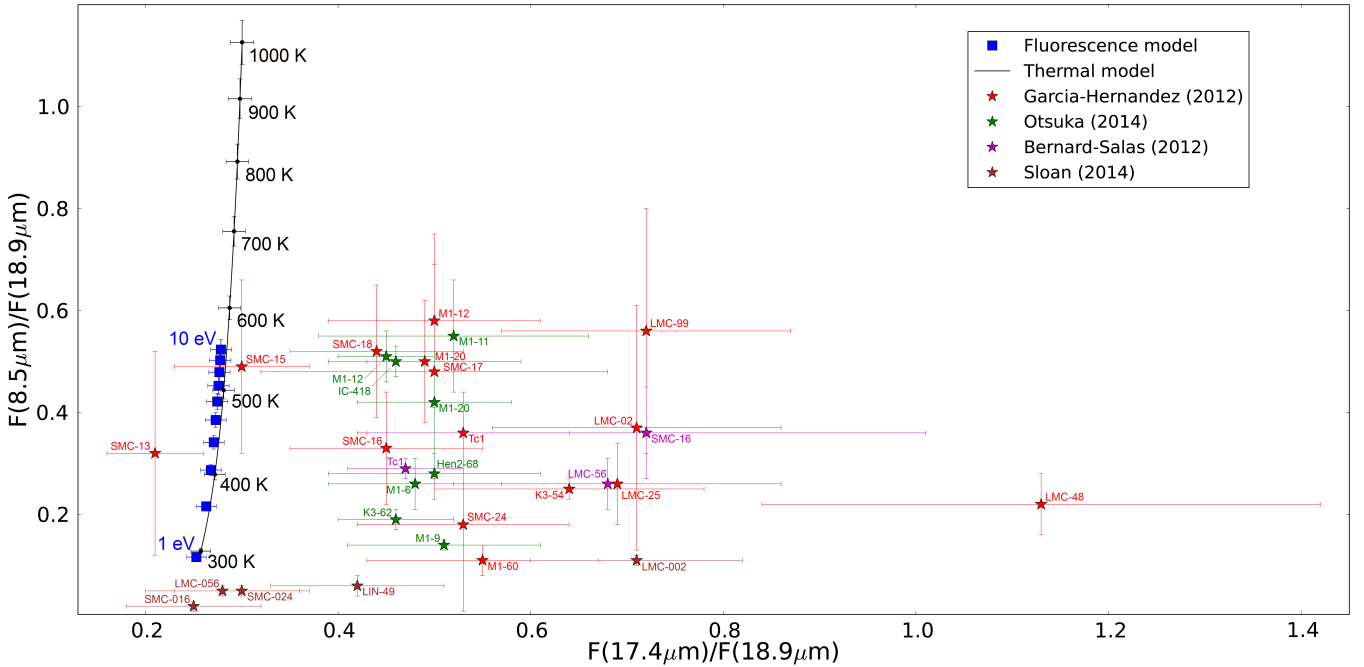

*Figure 5. Observational data versus modeled emission of $C_{60}$. Observed 8.5 μm/18.95 μm vs. 17.4 μm/18.95 μm emission flux ratios (stars) were taken from Bernard-Salas et al. (2012), García-Hernández et al. (2012b), Otsuka et al. (2014), Sellgren et al. (2010) and Sloan et al. (2014). The solid black curve represents the expected ratios for a thermal excitation, using our RIS, for temperatures ranging from 300 K to 1000 K. The blue squares represent the expected ratios for UV excitations using photon energies ranging from 1eV to 10 eV (see text).*

From the diagnostic plots presented in Figs. 5 and 6, it becomes clear that the observed 7.0μm/18.95 μm and 8.5 μm/18.95 μm flux ratios reported for most objects are well explained in terms of thermal or UV excitations, but that it is not possible to distinguish between thermal and UV excitation. On the other hand, the 17.4 μm/18.95 μm flux ratios show a large scatter from one object to the other, and most of them are in clear disagreement with the expected ratios of 0.25 - 0.30 which result from our RIS values, both for thermal and UV excitation. Using for instance the laboratory data from Iglesias-Groth et al.

(2011), modeled 17.4 μm/18.95 μm flux ratios of around 0.5 result, which are in better agreement with the bulk of the ratios inferred from the observations. A slightly wider range in the modeled 17.4 μm/18.95 μm ratios results if it is assumed the temperature dependence of the $C_{60}$ absorptivity reported by Iglesias-Groth et al. (2011) [3]. From their data, the modeled 17.4 μm/18.95 μm ratio for thermal excitation increases from 0.41 - 0.61, for temperatures ranging from 100 K to 400 K, and then starts to decrease again to values around 0.36, for temperatures around 1000 K.

The use of the laboratory data from Iglesias-Groth et al. (2011) does not alleviate the fundamental problem in the observed 17.4 μm/18.95 μm flux ratios. The large range in 17.4 μm/18.95 μm ratios of 0.2 - 1.2 cannot be consistently explained by one set of $C_{60}$ RIS values, as already shown by Bernard-Salas et al. (2012). In the following, we revisit the reported fluxes in the four $T_{1u}(i)$ modes of $C_{60}$, and briefly examine potential excitation scenarios other than thermal or fluorescent that may affect the observed emission.

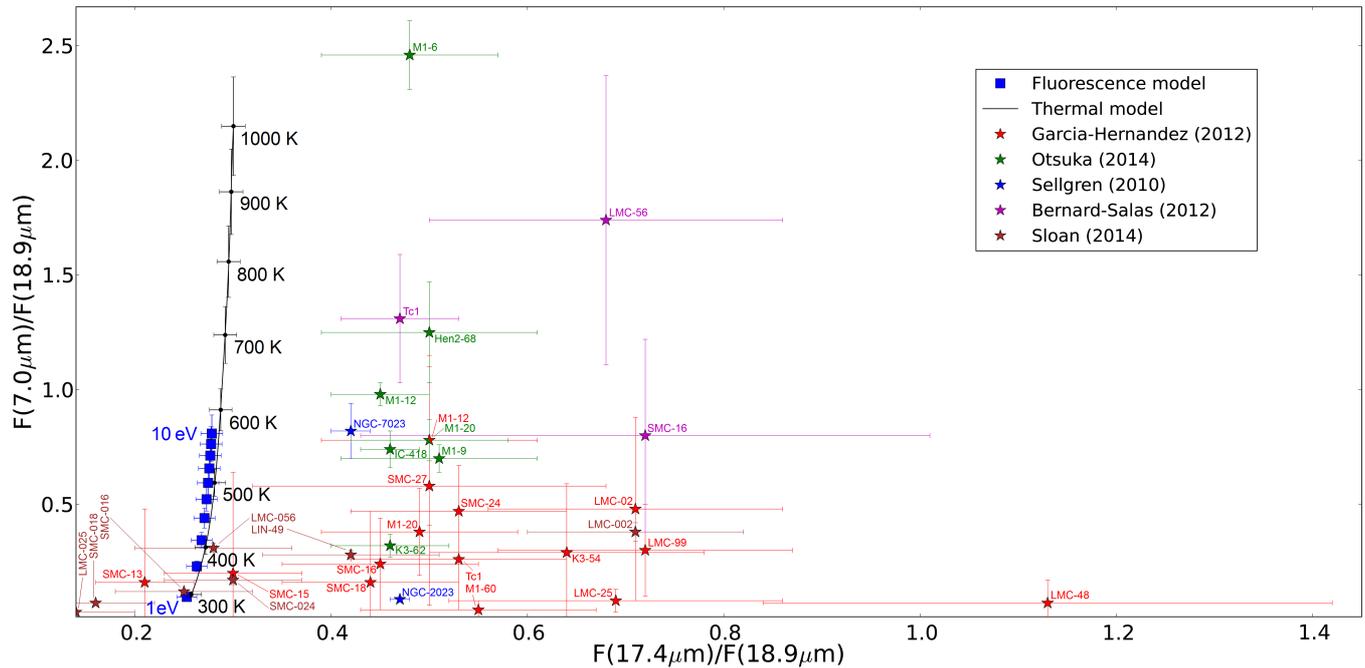

*Figure 6. As Fig. 5, but 7.0μm/18.95 μm vs. 17.4 μm/18.95 μm emission flux ratios are plotted.*

4.4 Discussion

In many sources, the observed emission near the four $T_{1u}(i)$ modes of $C_{60}$ is known to be contaminated by contributions from other fullerenes such as $C_{70}$ and $C_{60}^+$, PAH molecules, hydrogen recombination lines, and nebular ionized atomic and molecular emission including [ArII] or [SIII]. The wide range in values in the reported 17.4 μm/18.95 μm ratios may thus indicate that contributions from such emitters have been underestimated. For instance, both 17.4 μm and 18.95 μm emission features may contain significant contributions from $C_{70}$ which are unresolved from $C_{60}$ (e.g. Cami et al. 2010). Sloan et al. (2014) used the position and width of the observed 18.7 - 18.9 μm emission bands to scrutinize the presence of fullerenes in their spectra. While they claim positive detections of fullerenes in LIN 49, SMP LMC008 and SMP SMC 001, they questioned the presence of $C_{60}$ in SMP SMC 020, 027, and SMP LMC 099, which has been reported by García-Hernández et al. (2011a), mainly because the emission near 18.9 μm appears to be dominated by [SIII]. They also find that several sources which contain 21 μm emission features, show relatively strong and broad emission near 17.1 μm and near 6.9 μm. We also note that in many of the published spectra of Garcia-Hernandez et al. (2012b), the 17.4 μm band is barely present, while the 18.95 μm emission is strong in most sources and clearly detected. For instance, in the spectra of SMC-16, SMC-18, SMC-24, SMC-27, LMC-02, LMC-25, LMC-99, M1-20, and M1-12, the 17.4 µm emission is very faint indeed if not absent, yet Garcia-Hernandez et al. (2012b) report relatively large 17.4 µm/18.95 µm ratios (around 0.5) in most of those sources.

---

[3] We note that Iglesias-Groth et al. (2011) use very wide integration ranges when inferring $C_{60}$ absorptivities, and that their spectra seem to be affected by relatively strong scattering.

The laboratory spectra of $C_{60}^+$ show strong emission in its $E_{1u}(15)$, $E_{1u}(5)$, and ($A_{2u}(2)$ & $E_{1u}(3)$) modes, at wavelengths which occur near the $C_{60}$ $T_{1u}(4)$, $T_{1u}(2)$, and $T_{1u}(1)$ modes (e.g. Kern et al. 2013). In particular, Strelnikov et al. (2015) reported strong emission from $C_{60}^+$ near 18.95 µm. We might thus speculate that in several of the astrophysical sources, the observed emission arises from $C_{60}^+$ rather than from $C_{60}$. In the re-analysis of fullerene emission in carbon-rich circumstellar envelopes by Sloan et al. (2014), the authors pointed out that emission from 18.7 µm [SIII] contributes to the $C_{60}$ 18.9 µm feature, and that in particular in the LL2 low resolution Spitzer data (R = l/dl = 130), the [SIII] line is blended with the 18.9 µm $C_{60}$ emission. Otsuka et al. (2014), who detected $C_{60}$ in 11 out of a total sample of 338 PNe, estimated contributions of [ArII] to the 7 µm emission from the observed [ArIII] 8.99 µm emission line and theoretical [ArII]/[ArIII] ratios using photoionisation models. The authors note that the observed 8.5 µm emission may include contributions from the 8.6 µm emission from PAH molecules. We conclude that some of the reported fluxes in the four $T_{1u}(i)$ modes of $C_{60}$ are very uncertain indeed, and that they may not be used to infer physical conditions in the astrophysical environments.

The presence of fullerenes in hydrogen-rich environments has been challenged altogether by Duley & Hu (2012). Their laboratory spectra of hydrogenated amorphous carbon nanoparticles, which according to the authors do not contain $C_{60}$, show emission near the four $C_{60}$ features. The authors also detect emission at 16.4 µm in their samples, concluding that the emission in all five bands result from proto-fullerenes rather than from $C_{60}$, and suggest that $C_{60}$ is probably not present in astrophysical sources where PAH are detected. This hypothesis is in qualitative agreement with previous laboratory studies, which point out that fullerene formation might be enhanced in hydrogen-poor regions (e.g. Jäger et al. 2009). A top-down formation scenario of $C_{60}$ has also been proposed by Berné & Tielens (2012). The authors analyzed the existing data obtained in NGC 7023, where cold molecular material is irradiated by an intense UV flux from HD200775, and argue that the dehydrogenation of PAH into graphene would be followed by its conversion into $C_{60}$.

The confirmation of the presence of $C_{60}$ in some of the sources via optical absorption would help to settle the issue of the possible contributions from proto-fullerenes to the emission observed. Electronic transitions of $C_{60}$ occur at optical and UV wavelengths. Some weaker allowed transitions occur at 4024Å, 3980Å, and 3760 Å, yet the strongest electronic transitions of $C_{60}$ occur in the UV (e. g. Sassara et al. 2001). Optical absorption from $C_{60}$ was sought toward Tc-1 but surprisingly, no detection has been reported (García-Hernandez & Díaz-Luis 2013). A similar search has been conducted toward the Coronae Borealis (RCB) star DY Cen by García-Hernandez et al. (2012a), yet again, no $C_{60}$ absorption is seen. While the hydrogen-poor and carbon rich circumstellar envelopes of RCB stars have been suggested as formation sites of fullerenes, mid-infrared emission lines of $C_{60}$ has been detected in the regions of DY Cen and possibly V854 - surprisingly, the two least hydrogen-deficient RCB stars studied by Garcia-Hernandez et al. (2011b). Optical absorption from $C_{60}^+$ towards HD 200775 was sought as well and not detected (Berné 2013, unpublished). While all these findings may indicate that the bulk of the fullerene emission arises from spatial regions behind those tested in absorption, a comprehensive search of electronic transitions in fullerenes in astrophysical environments is warranted.

We briefly examine physical scenarios other than thermal or UV excitation that may affect the emission band ratios of the $C_{60}$ $T_{1u}$ modes. $C_{60}$ emission has been detected in a wide range of astrophysical environments, with widely varying physical conditions. We might speculate that formation of $C_{60}$ may very well imprint a population distribution among the vibrational modes different from that expected from thermal or UV excitation scenarios. The release of chemical energy in HAC nano-particles has been proposed to explain unidentified infrared emission features (Duley & Williams 2011), and a similar process may explain the large range observed in the 17.4 µm/18.95 µm ratios. Alternatively, it is interesting to note that under some astrophysical conditions, large molecules may suffer from an inverse internal conversion (IIC) of energy, as described by Léger et al. (1988). IIC occurs when the UV absorption is followed by rapid internal vibrational redistribution by recurrent or Poincaré fluorescence in the notion of Léger et al. (1988). A quantitative study of Poincaré fluorescence in the $C_{60}$ molecule is thus warranted.

## 5. CONCLUSIONS

We present a dedicated study for the determination of the RIS of the $T_{1u}$ bands of solid $C_{60}$, whose values reported so far show large discrepancies. We have discussed our experimental procedure for preparing the samples and the MIR spectroscopy of solid $C_{60}$ as a suitable technique for the determination of the RIS values to be used in astronomy. The samples prepared with our experimental method reveal high spectral stability, regardless of the morphological differences between the samples. We believe that this fact proves the validity of our RIS values (100: 25: 26: 40) for the interpretation of the astronomical observations.

Having a definite experimental knowledge of the RIS can help in the interpretation of the observational data in two possible ways:
1) When the observed emission is inconsistent with thermal and fluorescent mechanisms it can be concluded that unknown contributions are present, either from contamination or from other emission mechanisms, and therefore no physical conditions can be derived from the observational data.
2) In case that the contamination of the $C_{60}$ emission lines can be removed or neglected, improves the accuracy in the estimation of the physical conditions of the astrophysical source.

Using our RIS values, the observed emission band ratios 8.5 µm/18.95 µm and 7.0µm/18.95 µm are well explained in terms of a thermal or UV excitation of $C_{60}$. Both processes may however not be distinguishable from each other based on the astrophysical data. Our RIS values result in 17.4 µm/18.95 µm emission band ratios of 0.25 - 0.30 for temperatures of 260 K - 1000 K or photon energies of 1 - 10 eV. The observed 17.4 µm/18.95 µm emission band ratio range of 0.2 - 1.2 in the various astrophysical sources indicates that either additional emitters contribute to the observed emission near the $C_{60}$ $T_{1u}$ modes or that additional excitation mechanisms, such as chemo-luminescence from nascent $C_{60}$ or inverse internal conversion followed by Poincaré fluorescence, affect the population of the vibrational levels of $C_{60}$.

All of our samples show absorption near 6.5 µm, and we propose that neutral $C_{60}$ is the carrier of the unidentified emission band at 6.49 µm, which has been observed in fullerene-rich astrophysical objects.

## ACKNOWLEDGMENTS

We thank Prof. J. Cami for commenting on the manuscript. This research has been supported by the European Community's Seventh Framework Programme Marie-Curie IEF (Grant Agreement No. 274794).

APPENDIX A
EXPERIMENTAL DETAILS

In Figure 1(a) we show the schematic setup used for the preparation of the samples. $C_{60}$ was evaporated by laser desorption and deposited on potassium bromide substrates. The laser target consisted of $C_{60}$ powder (SES Research 99.5 %) placed at the bottom of a cylindrical crucible 3 mm deep and 6.5 mm wide. We used a Nd:YAG laser (Minilite - Continuum GmbH) operated at 532 nm. The maximum power density at the target was around 78 MW cm$^{-2}$ with pulses of 4 ns at a repetition rate of 10 Hz, giving an average power density of 3.1 W cm$^{-2}$. In order to increase the efficiency of condensation of the $C_{60}$ on the KBr, the substrate was placed such that it closes the entrance to the cylindrical crucible. The laser beam reached the $C_{60}$ enclosed in the crucible after having passed through the substrate. During the evaporation, the beam was moved across the $C_{60}$ target. The deposition took place in an argon atmosphere at reduced pressure (~ $10^{-2}$ mbar). As a result of the deposition, we obtained 6.5 mm wide spots of $C_{60}$ on the KBr substrates. The 36 samples analyzed in the present work were grown in this way, with deposition times varying from 1 to 5 minutes in order to produce samples of various thicknesses.

Spectra were collected in transmission mode using an FTIR spectrometer (Bruker VERTEX 80v) with a spectral resolution of 0.2 cm$^{-1}$ and with wavenumber and photometric accuracies of 0.01 cm$^{-1}$ and 0.1% respectively. Spectra were recorded from 400 to 6000 cm$^{-1}$ (25 to 1.7 μm), with separation of 0.5 cm$^{-1}$ between the data points.

In previous experiments, using the setup described above, we deposited $C_{60}$ on KBr substrates using different values of laser flux. This was done to explore possible side effects of the desorption on the MIR spectra of the samples. We observed that yellow-orange spots were deposited using a power density just above the minimum necessary for the evaporation, but black spots were produced when higher values were used (see Figure 1(b)). We found the threshold between both situations around 4 W cm$^{-2}$ (i.e. 100 MW cm$^{-2}$ during the pulse). MIR spectra of the spots deposited using a power density above this threshold showed a high variability in the RIS of the $T_{1u}$ bands, probably due to the formation of amorphous carbon (Section 3.2).

As described above, all the samples used in our analysis were produced with a laser flux below this threshold. The only requirement for the morphology of the resulting samples was to be in the optimum range of thickness for MIR transmission spectroscopy, allowing large differences in thickness from sample to sample and also at different locations throughout every sample, with less material in the areas of the substrate crossed by the laser beam during the deposition.

In order to obtain additional information on the purity of the samples produced, VIS Raman spectroscopy of the $C_{60}$ films was performed. The spectra were recorded using the Micro-Raman spectrometer Labram II (Fa. Dilor) equipped with a microscope and an internal He-Ne laser that emits photons at a wavelength of 632.8 nm. The measurements were carried out in air at room temperature. Exposure times were kept under 20 seconds to avoid sample damage. A possible damage or processing of the sample could be excluded by monitoring the surface of the sample before and after the Raman measurements using a visual imaging system.

APPENDIX B
WHY SOLID $C_{60}$ AT ROOM TEMPERATURE?

B.1 Aggregated $C_{60}$ versus isolated $C_{60}$

Observational data does not allow an unambiguous conclusion on the state of aggregation of the astrophysical $C_{60}$. The position of the $T_{1u}(2)$ band around 570 cm$^{-1}$ (17.5 μm) might indicate that the astrophysical $C_{60}$ is in the gas phase, because it is close to the position reported by Frum et al. (1991) and Nemes et al. (1994) from laboratory experiments at 1000 K. On the other hand, frequencies close to 578

cm$^{-1}$ (17.3 µm) would indicate solid state $C_{60}$ (Evans et al. 2012). Unfortunately, experimental uncertainties and the dependence of the $T_{1u}(2)$ position with the temperature does not allow to be conclusive on this point. Besides, if we consider the spectrum measured in cryogenic matrices of noble gases as an approximation to the gas phase spectrum of $C_{60}$, we notice that the position of $T_{1u}(2)$ in Ar and Kr reported by Haufler et al. (1990), and in Ne by Kern et al. (2013) around 578 cm$^{-1}$ (17.3 µm) is inconsistent with the hypothesis exposed above. MIR absorption spectroscopy of gas phase $C_{60}$ could contribute to clarify this point but, unfortunately, attempts to obtain a spectrum of $C_{60}$ in supersonic jets have failed due to its low vapor pressure and the inefficient vibrational cooling of the $C_{60}$ in jets (Stewart et al. 2013).

It is well established that $C_{60}$ forms a molecular solid, in which intermolecular interactions are much weaker than the intramolecular covalent forces. Therefore, it can be assumed with a great degree of accuracy that the $C_{60}$ molecules are essentially free rotors in their lattice sites, at least above the transition temperature of 260 K (Menendez & Page 2000).

On the other hand, the use of $C_{60}$ isolated in cryogenic matrices presents the problem of band splitting at low temperature, which precludes an accurate evaluation of the RIS. Splitting does not occur or is negligible at room temperature. The cause of the splitting is still under debate, some authors attribute the splitting to the isotopic substitution (Homes et al. 1994; Homes et al. 1995), and others to factor-group splitting (Kornienko et al. 2010).

B.2 Range of temperatures required for the analysis of the observational data.

Regarding the adequate temperature range to carry out the experiments, let us first consider that $C_{60}$ undergoes a first-order phase transition at 260 K (David et al. 1992; Heiney et al. 1991), which divides the range of temperatures into above and below 260 K.

Changes in the integrated absorptivity (called band strength throughout the paper) of the $T_{1u}$ bands with temperature have been reported but most of the studies are mainly focused in the characterization of the phase changes (Babu & Seehra 1992; Chase et al. 1992; David et al. 1992; Heiney et al. 1991; Narasimhan et al. 1992). Numerical dependence within our temperature range can be found in Iglesias-Groth et al. (2011) and also can be worked out from the data published by (Graja 1997; Graja & Swietlik 1995). Unfortunately, both sources report opposite dependences of $T_{1u}(2)/T_{1u}(1)$ with temperature. Regarding the negligible intermolecular interaction of solid $C_{60}$ above 260 K, the most probable scenario within this range is a weak dependence of the RIS on the temperature.

Thus, we believe that the experiments in the solid phase at room temperature presented here are the best choice to obtain the RIS to be used in the interpretation of the astronomical observations, even if the astrophysical $C_{60}$ is in gas phase. We present some more discussion on the nature of solid $C_{60}$ in Section 3.1.3.

TABLES

Table 1. Bands of the 7 μm region discussed in the text.

| Experiment | | | Theory | |
|---|---|---|---|---|
| Band position | | FWHM | Possible assignment | Position [a] (cm$^{-1}$) |
| (cm$^{-1}$) | (μm) | (cm$^{-1}$) | | |
| 1424 [b] | 7.02 | 3 [b] | $H_g(7)$ [c] | 1425 |
| 1429.3 ± 0.1 | 7.00 | 5.7 ± 0.2 | $T_{1u}(4)$ | 1429 |
| 1444 ± 2 | 6.9 | 6 ± 1 | $H_g(4) \otimes H_u(3)$ | 1440 |
| | | | $H_g(5) \otimes T_{3u}(1)$ | 1442 |
| | | | $G_g(1) \otimes G_u(4)$ | 1446 |
| 1451 ± 2 | 6.9 | 7 ± 2 | $H_g(3) \otimes H_u(4)$ | 1452 |
| | | | $H_g(5) \otimes G_u(1)$ | 1452 |
| | | | $H_g(1) \otimes T_{1u}(3)$ | 1454 |
| 1462 ± 1 [d] | 6.8 | 10 ± 2 | $G_g(1) \otimes T_{3u}(3)$ | 1458 |
| | | | $H_g(3) \otimes T_{3u}(2)$ | 1462 |
| | | | $T_{3g}(3) \otimes H_u(3)$ | 1464 |
| 1539.4 ± 0.2 | 6.496 | 3.8 ± 0.6 | $H_g(4) \otimes G_u(2)$ | 1536 |
| | | | $T_{3g}(3) \otimes H_u(4)$ | 1539 |
| | | | $G_g(2) \otimes T_{3u}(3)$ | 1540 |

[a] Combination modes which might correspond to the bands according to the first-order frequencies listed by Menendez & Page (2000).
[b] Fixed value of position and FWHM used for fitting.
[c] Raman-active mode, reported to be infrared-active by Martin et al. (1994).
[d] In some of the spectra, this band appears to be formed by two components at 1457.5 and 1463.5 cm$^{-1}$ (6.86 and 6.83 μm) with FWHMs of roughly 6 cm$^{-1}$.

Table 2. Relative intrinsic strength of the main $C_{60}$ bands.

| Contributions included at 7 μm | MIR bands Relative intrinsic strength (%) | | | |
|---|---|---|---|---|
| | 18.9 μm | 17.4 μm | 8.5 μm | 7.0 μm |
| $T_{1u}(4)$ | 100 | 25 ± 1 | 26 ± 1 | 38 ± 4 |
| $T_{1u}(4) + H_g(7)$ | 100 | 25 ± 1 | 26 ± 1 | 40 ± 4 |
| 7 μm structure | 100 | 25 ± 1 | 26 ± 1 | 59 ± 8 |

Table 3. Experimental parameters of the bands assigned to the $T_{1u}$ modes.

| Band | Experimental Position [a] | | FWHM |
|---|---|---|---|
| | (cm$^{-1}$) | (μm) | (cm$^{-1}$) |
| $T_{1u}(1)$ | 527.00 ± 0.06 | 18.98 | 3.1 ± 0.2 |
| $T_{1u}(2)$ | 576.38 ± 0.12 | 17.35 | 3.0 ± 0.3 |
| $T_{1u}(3)$ | 1182.90 ± 0.09 | 8.45 | 3.8 ± 0.2 |
| $T_{1u}(4)$ | 1429.33 ± 0.10 | 7.00 | 5.7 ± 0.2 |

[a] Mean and standard deviation obtained from the fitting.